\documentclass[letter]{ptptex}
\usepackage{graphicx}
\usepackage{wrapft}

\title{
Hyperons in Nuclear Matter%
}

\author{
Tsuyoshi \textsc{Miyatsu} and Koichi \textsc{Saito}%
}

\inst{
Department of Physics, Faculty of Science and Technology \\
Tokyo University of Science, Noda, Chiba 278-8510, Japan
}

\abst{
The chiral version of the QMC model, in which the effect of gluon and pion exchanges is included 
self-consistently, is applied to the hyperons in a nuclear medium. 
The hyperfine interaction due to the gluon exchange plays an important role in the 
in-medium baryon spectra, while the pion-cloud effect is relatively small. 
At the quark mean-field level, the $\Lambda$ feels more attractive force than the $\Sigma$ or $\Xi$ in matter. 
}

\begin{document}

\maketitle


The study of $\Lambda$ hypernuclei has a long and impressive history.\cite{history,HT}  
In particular, it is well known that the spin-orbit force for $\Lambda$ hypernuclei is very weak,\cite{s-o} and systematic 
studies of the energy levels of $\Lambda$ hypernuclei provide us with significant information on the $\Lambda$-nucleon (N) 
interaction, the deep nuclear interior and a possible manifestation of the quark degrees of freedom in a nuclear medium.  
Current studies of $\Lambda$ hypernuclei at JLab are also very interesting.\cite{HT} 

It seems likely that the situation in $\Sigma$ hypernuclei is, however, quite different,\cite{noumi} 
because, despite extensive searches, there 
is only one experimental evidence for the $\Sigma$ hypernuclei, namely $^4_\Sigma$He.\cite{outa} Thus, it is a challenging 
experiment to produce $\Sigma$ hypernuclei or doubly strange ($\Xi$) nuclei.\cite{hiyama1}  We eagerly await forthcoming 
studies of hypernuclei with new facilities at J-PARC and GSI-FAIR, and they will give us important knowledge of  
the hyperon(Y)-N interaction in nuclei.\cite{hiyama2} 

In our earlier works we addressed the question of whether quarks play an important role in finite nuclei, 
using the quark-meson coupling (QMC) model.\cite{ppnp} Starting at the quark level, the QMC model was created to provide 
insight into the structure of nuclear matter.\cite{guichon} 
The model can naturally lead to an explanation of 
the small spin-orbit force in the $\Lambda$ hypernuclei.\cite{hyper1}  Guichon {\it et al.} recently calculated the energy levels of 
hypernuclei using the latest development of the QMC model,\cite{hyper2} and found that the hyperfine interaction due to the one-gluon 
exchange (OGE) is very important to describe the properties of $\Sigma$ or $\Xi$ hypernuclei. 

In this work we apply the chiral quark-meson coupling (CQMC) model\cite{cqmc} to study the properties of hyperons 
($\Lambda$, $\Sigma$, $\Xi$) in a nuclear matter. The CQMC model is an extended version of the QMC model to incorporate chiral symmetry, 
and it is based on the volume coupling version of the cloudy bag model (CBM).\cite{tony} After linearizing the pion cloud surrounding a  
nucleon, the Lagrangian density is given by 
${\cal L}_{CBM} = {\cal L}_{BAG} + {\cal L}_{\pi} + {\cal L}_{g} + {\cal L}_{int}$, 
where the usual, bag Lagrangian density is\cite{mit}  
\begin{equation}
{\cal L}_{BAG} = \left[ {\bar \psi} ( i\gamma_\mu \partial^\mu -m_i) \psi -B \right] \theta_V - \frac{1}{2}{\bar \psi} \psi 
\delta_S , 
\end{equation}
with $\psi$ the quark field, $m_{i=0(s)}$ the current $u, d$ ($s$) quark mass, $B$ the bag constant, 
$\theta_V$ the step function for the bag and $\delta_S$ the surface $\delta$-function. The interaction Lagrangian density is\cite{cqmc} 
\begin{equation}
{\cal L}_{int} = {\bar \psi} \left[ i \frac{m_0}{f_\pi}\gamma_5 {\vec \tau} \cdot {\vec \phi} 
+ \frac{1}{2f_\pi} \gamma_\mu \gamma_5 {\vec \tau} \cdot (\partial^\mu {\vec \phi}) 
+ \frac{g}{2} \gamma_\mu {\vec \lambda} \cdot {\vec A}^\mu  \right] \psi \, \theta_V  , \label{int-lag}
\end{equation}
with ${\vec \lambda}$ the SU(3) color generators, 
$f_\pi$ ($=93$ MeV) the pion decay constant and $g$ the quark-gluon coupling constant. 
Here the pion field, ${\vec \phi}$, interacts with the {\it light} quark through the pseudovector (pv) and 
pseudoscalar (ps) couplings.  
The strength of the ps coupling is ${\cal O}(m_0 / f_\pi)$, which explicitly shows the breaking scale of chiral symmetry.  
The free pion field and the kinetic energy of the gluon field, ${\vec A}^\mu$, are, respectively, described by ${\cal L}_{\pi}$ and 
${\cal L}_{g}$.  

\begin{figure}
\centerline{\includegraphics[width=10 cm,height=10 cm]{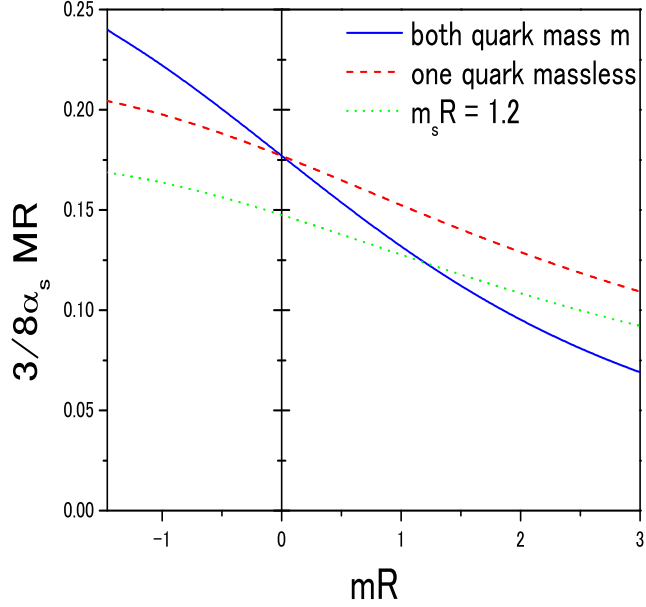}}
\caption{$M_{ij}$ as a function of the quark mass ($\delta =mR$). The green (dotted) curve presents the case where 
one quark is strange and $m_sR$ is fixed to be 1.2.}
\label{fig:m-ij}
\end{figure}

\begin{wraptable}{l}{\halftext}
\caption{Values of $a_{00}$, $a_{0s}$ and $a_{ss}$ for $N$, $\Delta$, $\Lambda$, $\Sigma$ and $\Xi$. }
\label{table:g-mag}
\begin{center}
\begin{tabular}{cccc} \hline \hline
 & $a_{00}$ & $a_{0s}$ & $a_{ss}$ \\ \hline
$N$ & $-2$ & $0$ & $0$ \\
$\Delta$ & $2$ & $0$ & $0$ \\ 
$\Lambda$ & $-2$ & $0$ & $0$ \\
$\Sigma$ & $2/3$ & $-8/3$ & $0$ \\ 
$\Xi$ & $0$ & $-8/3$ & $2/3$ \\ \hline
\end{tabular}
\end{center}
\end{wraptable}
The OGE contribution to the hadron mass can be split into the electric ($\Delta E_G^{el}$) and magnetic 
($\Delta E_G^{mg}$) parts, and the former is negligibly small because of the color-charge neutrality of hadron.\cite{mit,inoue} 
Of importance is the color magnetic interaction between two quarks: 
\begin{equation}
\Delta E_G^{mg} = \frac{\alpha_s}{R}[ a_{00}M_{00} + a_{0s}M_{0s} + a_{ss}M_{ss} ] , \label{g-mag}
\end{equation}
where $R$ is the bag radius, $\alpha_s = g^2/4\pi$ and the subscript $0 (s)$ of $a_{ij}$ or $M_{ij}$ denotes the light (strange) quark. 
The coefficient $a_{ij}$ is the expectation value of the operator, $\sum_{i>j}({\vec \lambda}{\vec \sigma})_i \cdot 
({\vec \lambda}{\vec \sigma})_j$, with respect to the SU(6) color-spin wavefunction\cite{mit}, and it is  
presented in Table~\ref{table:g-mag}. Then, $M_{ij}$ is given in terms of the quark magnetic moment $\mu$ as 
\begin{equation}
M_{ij}(m_i, m_j, R) = \frac{3}{R^3} \mu(m_i, R) \mu(m_j, R) I(\delta_i, \delta_j) , \label{qmag}
\end{equation}
where $\delta_{i} = m_{i}R$ and $I$ is also a function of the magnetic moment.\cite{mit}  The dependence of $M_{ij}$ on 
the quark mass is shown in Fig.~\ref{fig:m-ij}. 
%
%
%

This magnetic force differently contributes 
to the $\Lambda$ and $\Sigma$ hyperons, because the $u$ and $d$ quarks in $\Lambda$ ($\Sigma$) are in the 
spin-singlet (triplet) state. In fact, the magnetic contributions to $\Lambda$ and $\Sigma$ are, respectively, proportional to 
$-3M_{00}$ and $M_{00}-4M_{0s}$.  These contributions lower both masses of $\Lambda$ and $\Sigma$, and the mass reduction in 
$\Lambda$ is much larger than in $\Sigma$.  This is because $M_{00} > M_{0s} > M_{ss}$ (see Fig.~\ref{fig:m-ij}).  
In $\Xi$, the mass reduction is given in terms of $-4M_{0s}+M_{ss}$, and thus it is larger than the $\Sigma$-mass reduction. 

We next calculate the second-order energy correction caused by the pion-quark interaction 
in Eq.(\ref{int-lag}).  The energy shift due to the interaction is calculated by the Hubbard's prescription.\cite{cqmc,inoue,saito} 
%
%
Since our aim is to see how the gluon and pion exchanges affect the difference between the hyperon masses in free space and in a medium, 
we here calculate a sum of the direct and exchange contributions, namely the Hartree-Fock (HF) contribution, $\Delta E_{HF}$.

%
%
%
Then, the HF result for the pv coupling is calculated as\cite{cqmc}
\begin{eqnarray}
\Delta E_{HF}^{pv} &=& 
-\frac{[{\vec \sigma} {\vec t}\, ]_{\pi} N^4}{12\pi^2 x^3 (f_\pi R)^2 R}   
\int_0^x d\rho_1 \rho_1^2 \int_0^x d\rho_2 \rho_2^2 \int_0^\infty \frac{dt \, t^4}{t^2+y^2} 
\biggl[ A(\rho_1) A(\rho_2) j_0(t\rho_1) j_0(t\rho_2)  \nonumber \\
&+& \frac{1}{3} 
\left\{ A(\rho_1) B(\rho_2) j_0(t\rho_1) \left( j_0(t\rho_2) -2 j_2(t\rho_2) \right) 
+ (1 \leftrightarrow  2)  \right\} \nonumber \\
&+& \frac{1}{9} B(\rho_1) B(\rho_2) \left( j_0(t\rho_1) -2 j_2(t\rho_1) \right) 
\left( j_0(t\rho_2) -2 j_2(t\rho_2) \right) \biggr] ,  \label{hfpv1}
\end{eqnarray}
with $N$ the normalization constant for the quark wave function, $x$ the lowest quark eigenvalue in the bag, 
$y=m_\pi R/x$ ($m_\pi = 138$ MeV the pion mass), $A(\rho) = j_0^2(\rho) - \beta^2 j_1^2(\rho)$, $B(\rho) = 2 \beta^2 j_1^2(\rho)$, 
$\beta = x/(\lambda + \delta)$ and $\lambda^2 = x^2 + \delta^2$.  Here, $[{\vec \sigma} {\vec t}\, ]_{\pi}$ is 
the spin-isospin matrix element, and for the N or delta ($\Delta$) it is given by\cite{chin}
\begin{equation}
[{\vec \sigma} {\vec t}\, ]_{\pi} = 
\sum_{i \neq i^\prime\in N, \Delta} \langle i| {\vec \sigma} {\vec t} |i^\prime \rangle \cdot \langle i^\prime| {\vec \sigma} 
{\vec t} |i \rangle 
= 9-S(S+1) - I(I+1) ,  \label{me1}
\end{equation}
where the index ($i, i^\prime$) runs over the spin and isospin, and 
$S \, (I)$ is the total spin (isospin) of N or $\Delta$. 

The contribution from the ps interaction or the interference between pv and ps interactions can 
be calculated in the similar manner. They are, however, very small, compared with the pv contribution.\cite{cqmc} 

The pion correction, $\Delta E_{HF}^{pv}$, diverges like $- 1/R^3$ as $R \to 0$, and thus the bag collapses. 
Because the pion has, however, a finite size, the $q{\bar q}$ structure of the pion is essentially important when the bag 
radius is very small. In Ref.\cite{saito}, a phenomenological, non-local interaction was studied 
to settle this collapse at $R \sim 0$. 
The effect of the $q{\bar q}$ structure of pion can eventually be described by a form factor at the vertex of the 
quark-pion interaction. When the charge radius of the pion is about $0.56$ fm\cite{dally}, 
the form factor is estimated as $F_{q\pi}(R) = [1 + 1.3\times (b / R)^2 ]^{-3/2}$ with $b = 0.46$ fm (see Fig.2 in Ref.\cite{saito}).  
Using this form factor, one can regularize the pion contribution at $R = 0$.\cite{cqmc} 

The pion corrections to the N, $\Delta$, $\Lambda$, $\Sigma$ and $\Xi$ are in the ratio $15 : 3 : 9 : 1 : 0$.\cite{inoue,saito}  
Thus the pion does not contribute to the $\Xi$ mass, while it lowers the $\Sigma$ mass.  
It is, however, important to notice that 
the pion contribution to the $\Sigma$ mass is not large and its amount is about $1/15$ of that to the N mass (see below).  

The mass of the N or $\Delta$ in free space is given by a sum of the usual 
bag energy\cite{mit} and the corrections due to the pion and gluon exchanges. In the present calculation, 
we fix the current $u$, $d$ quark mass, $m_0 = 5$ MeV, because the dependence of the baryon mass on $m_0$ is not strong.\cite{ppnp} 
Thus there are four parameters: $B$, $z_N$, $z_\Delta$ and $\alpha_s$. 
Since we can expect that the usual $z$ parameter for the N, which accounts for the center of mass correction, is not much different from 
that for the $\Delta$, we here 
choose $z_0 = z_N = z_\Delta$. Then, the bag constant, $B$, and $z_0$ are determined so as to fit 
the free nucleon mass, $M_N (= 939$ MeV), with its bag radius, $R_N = 0.8$ fm. 
(The following numerical results weakly depend on $R_N$.\cite{ppnp})
The remaining parameter, $\alpha_s$, is chosen so as to yield the correct mass difference between $M_N$ and the $\Delta$ mass, 
$M_\Delta (=1232$ MeV).  

\begin{wraptable}{l}{\halftext}
\caption{Bag parameters for $m_0 = 5$ MeV and $R_N = 0.8$ fm. The bag constant and the strange quark mass are in MeV.}
\label{table:bagparam}
\begin{center}
\begin{tabular}{ccccc} \hline \hline
case & $B^{1/4}$ & $z_0$ & $\alpha_s$ & $m_s$ \\ \hline
$1$ & 168.8 & 2.476 & 0.367 & 275.6 \\
$2$ & 169.9 & 2.671 & 0.443 & 286.6 \\ \hline
\end{tabular}
\end{center}
\end{wraptable}
In this paper, we study two cases where the pion contribution is included 
(case $1$) or not included (case $2$). In Table~\ref{table:bagparam}, we present the bag parameters for 
the case $1$ or $2$. 
In addition to the bag parameters, we use the strange quark mass, $m_s$, to fit the $\Omega$ mass ($M_\Omega = 1672$ MeV). 

The N-$\Delta$ mass difference (about $300$ MeV) is mainly reproduced by  
the OGE contribution (about $240$ MeV). In contrast, the pion-exchange contribution is 
about $60$ MeV, which is near the upper limit allowed from lattice QCD constraints\cite{young}.   
The bag radius of the $\Delta$ is calculated to be $R_\Delta = 0.88$ fm (for the case 1). 

We also calculate the mass of hyperon ($\Lambda$, $\Sigma$, or $\Xi$).  In this calculation, we take a different 
$z$ parameter for each hyperon and fit the calculated mass to the observed value in free space.  We then find 
$z_\Lambda = 2.488 \ (2.601)$, $z_\Sigma = 2.423 \ (2.418)$ and $z_\Xi = 2.507 \ (2.498)$ for the case 1 (2). 

To describe a nuclear matter, we need the intermediate attractive and short-range repulsive nuclear forces. 
As in the QMC model,\cite{ppnp} it is achieved by introducing the $\sigma$ and $\omega$ mesons. 
However, the present $\sigma$ meson is assumed to be chirally singlet and {\em not} the chiral partner of the $\pi$ meson.  
This $\sigma$ represents, in some way, the exchange of two pions in the iso-scalar N-N interaction.\footnote{
As in chiral perturbation theory,\cite{cpt} instead of assuming the $\sigma$, it may be possible to obtain the intermediate attractive 
force by calculating the two-pion exchange diagrams between two nucleons. 
Since such calculation is, however, very hard, we leave it as the future study.\cite{cqmc} 
} 

Now let us start from the Lagrangian density for the CQMC model:\cite{cqmc} ${\cal L}_{CQMC} = {\cal L}_{CBM} + {\cal L}_{\sigma \omega}$, 
where 
\begin{equation}
{\cal L}_{\sigma \omega} = {\bar \psi} \left[ g_\sigma^q \sigma - g_\omega^q \gamma_0 \omega \right] \psi \, \theta_V  
- \frac{1}{2} m_\sigma^2 \sigma^2 + \frac{1}{2} m_\omega^2 \omega^2 ,  \label{qmc-lag2}
\end{equation}
with $g_\sigma^q (g_\omega^q)$ the $\sigma (\omega)$-quark coupling constant, $m_\sigma (m_\omega)$ the meson mass 
and $\sigma (\omega)$ the mean-field value of the $\sigma$ ($\omega$) meson.  We here assume that a strange quark does not couple to the 
mesons. In an iso-symmetric nuclear matter, the total energy per nucleon is then given by
\begin{equation}
E_{tot} = \frac{4}{(2\pi)^3 \rho_B} \int^{k_F} d{\bf k} \sqrt{{\bf k}^2+M_N^{*2}} + 3g_\omega^q \omega 
+ \frac{1}{2}( m_\sigma^2 \sigma^2 - m_\omega^2 \omega^2 ) ,  \label{tot}
\end{equation}
with $\rho_B = 2k_F^3/3\pi^2$ the baryon density and $M_N^{*}$ the effective nucleon mass calculated by the CBM. 
(Hereafter, the asterisk $^*$ denotes the quantity in matter.)

The attractive force due to the $\sigma$ exchange leads to the modification of the in-medium quark mass as 
$m^* = m - g_\sigma^q \sigma$, and it changes the quark wave function. 
This modification then generates the effective nucleon mass in matter. 
Because the change of the quark wave function varies the source of the $\sigma$ field, 
we have to solve the coupled, nonlinear equations for the nuclear matter self-consistently (for details, see Refs.\cite{ppnp}).  

%
\begin{table}
\caption{Coupling constants and calculated properties for symmetric nuclear matter at $\rho_0$. 
The last three columns show the relative changes (from their values at zero density) of the nucleon bag radius, the lowest eigenvalue for 
the quark and the root-mean-square radius of the nucleon calculated with 
the quark wave function. The nucleon mass and the nuclear incompressibility, $K$, are in MeV.}
\label{table:cc}
\begin{center}
\begin{tabular}{ccccccccc} \hline \hline
case & $g_\sigma^2/4\pi$ & $g_\omega^2/4\pi$ & $M_N^*$ & $K$ & $\delta R_N^*/R_N$ & $\delta x^*/x$ 
& $\delta r^*/r$ \\ \hline
$1$ & 5.67 & 6.92 & 717 & 337 & -0.01 & -0.19 & 0.03 \\
$2$ & 4.55 & 5.95 & 740 & 299 & -0.02 & -0.15 & 0.02 \\ \hline
\end{tabular}
\end{center}
\end{table}
The numerical result is presented in Table~\ref{table:cc}.  We take $m_\sigma = 550$ MeV and $m_\omega = 783$ MeV.  
The $\sigma$-N and $\omega$-N coupling constants, $g_\sigma (= 3g_\sigma^q S(\sigma=0)$) and $g_\omega (= 3g_\omega^q$), 
are determined so as to fit 
the nuclear saturation condition ($E_{tot} - M_N = -15.7$ MeV) at normal nuclear density $\rho_0 (= 0.15$ fm$^{-3}$). 
Here $S(\sigma)$ is the quark scalar density calculated with the quark wave function.\cite{ppnp,guichon,cqmc} 
%
%

In the present model, although the bag radius slightly shrinks, the root-mean-square (rms) radius 
swells by about $3$\% at $\rho_0$.  It is caused by the attractive nuclear force and the amount of swelling is 
well within the experimental constraint.\cite{electron} The quark eigenvalue, $x$, decreases by about $20$\%, which leads to 
the smaller in-medium nucleon mass than in the usual QMC model.  

\begin{wrapfigure}{l}{6.6cm}
\centerline{\includegraphics[width=6.5 cm]{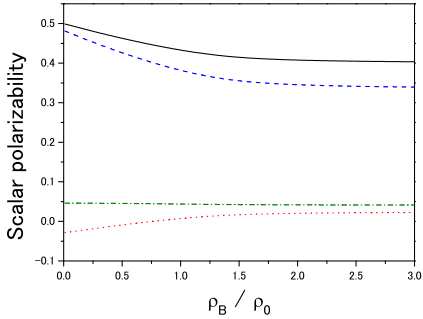}}
\caption{Scalar polarizability (for the case 1). The red (dotted) curve presents 
the pion contribution, while the green (dot-dashed) curve shows the gluon one.  
The scalar density $S(\sigma)$ is presented by the blue (dashed) curve. The total sum is shown 
by the black (solid) curve. 
}
\label{fig:spol}
\end{wrapfigure}
In the QMC model, of importance is the scalar polarizability, $P$, which describes the response of a quark to 
the external scalar field.\cite{ppnp}  It is defined by the derivative of $M_N^*$ with respect to the $\sigma$ field, 
$P = - \partial M_N^*(\sigma) / \partial (3g_\sigma^q \sigma)$, and it 
consists of the quark scalar density, $S(\sigma)$, and the contributions from the gluon and pion exchanges 
(see Fig.~\ref{fig:spol}). 
We find that even in the CQMC model the scalar polarizability decreases with increasing $\rho_B$. 
Because of this reduction in matter, the present model can provide a much smaller value 
of the nuclear incompressibility than in Quantum Hadrodynamics ($K_{QHD} \sim 550$ MeV). 
The calculated incompressibility is close to the observed value ($K_{exp} \sim 200-300$ MeV).

How is the hyperon mass modified in matter?  If the effect of the gluon and pion exchanges is not included, the QMC model 
simply predicts the scaling law for the hadron ($H$) mass,\cite{scaling} that is  
$\delta M_\Delta^*/\delta M_N^* : \delta M_\Lambda^*/\delta M_N^* : \delta M_\Sigma^*/\delta M_N^* : \delta M_\Xi^*/\delta M_N^* 
= 1 : 2/3 : 2/3 : 1/3$, where $\delta M_H^* = M_H - M_H^*$. The factor $1$, $2/3$ and $1/3$ come from the ratio of the number of 
light quarks in $H$ to that in the N. This means that the hadron mass is practically determined by only the number of light quarks, which 
{\it feel} the common scalar field generated by surrounding nucleons in matter. 

In contrast, when the effect of gluon and pion is considered, the situation is different.  It provides the hyperfine 
splitting in the hadron spectra. 
Because the light quark feels the attractive force due to the $\sigma$ exchange, it is {\it more} relativistic than in free space.  
The quark magnetic moment is thus enhanced in matter.\cite{qmag} Such modification leads to the enhancement of the color magnetic 
interaction in matter.  We can clearly see it in Fig.~\ref{fig:m-ij}.  
Note that $M_{ss}$ is not changed in matter because the strange quark does not feel any attraction. 
It is thus expected that $\delta M_\Lambda^* > \delta M_\Sigma^*$ in matter. 
We here recall that the $\Xi$-mass reduction due to the OGE is larger than the $\Sigma$ one, 
while the pion correction reduces the $\Sigma$ mass but does not affect the $\Xi$ mass.  

\begin{figure}[htb]
 \parbox{\halftext}{\centerline{\includegraphics[width=7.5 cm]{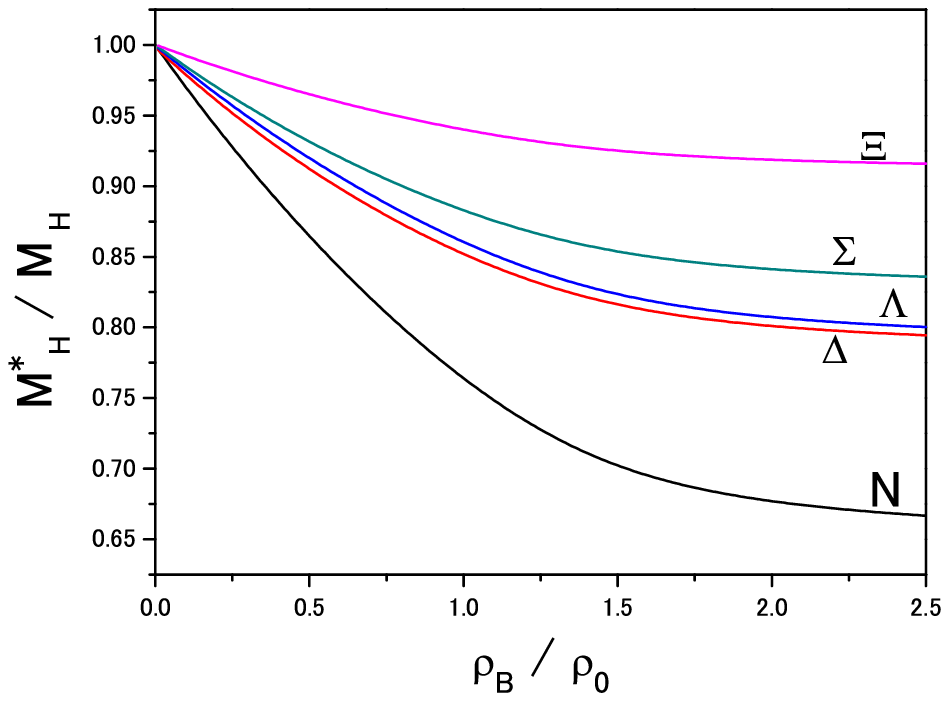}}
 \caption{Ratio of the baryon mass in matter to the free one (for the case 1). } \label{fig:rmas}
}
 \hfill
 \parbox{\halftext}{\centerline{\includegraphics[width=7.5 cm]{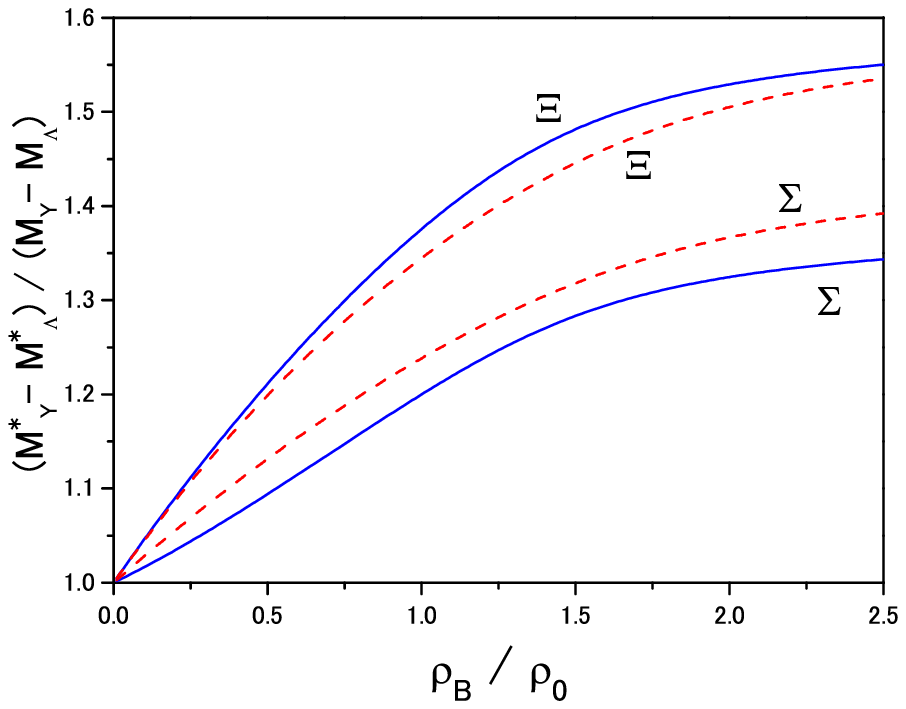}}
 \caption{Ratio of the mass difference, $M_Y^*-M_\Lambda^*$, in matter to that in free space ($Y=\Sigma$ or $\Xi$). 
 The blue (solid) curve is for the case 1, while the red (dashed) curve is for the case 2. } \label{fig:hypmas}
}
\end{figure}

Now we are in a position to show our main results. In Fig.~\ref{fig:rmas}, we present the baryon mass in a nuclear 
matter.  
The scaling is apparently violated. The $\Sigma$ and $\Lambda$ masses are split up in matter. 
The N-$\Delta$ mass difference is very enhanced, because of the color magnetic 
interaction $M_{00}$.  As seen in Table~\ref{table:g-mag} and Fig.~\ref{fig:m-ij}, it increases $M_\Delta^*$ but reduces $M_N^*$. 
The pion-cloud effect is relatively minor in the baryon spectra.

In Fig.~\ref{fig:hypmas}, we show the ratio of the mass difference, $M_Y^*-M_\Lambda^*$ ($Y=\Sigma$ or $\Xi$), in matter to 
that in free space. Comparing with the mass difference in free space, the ratio increases as $\rho_B$ goes up. 
This means that, in matter, the $\Sigma$ or $\Xi$ feels {\it less} attractive force than the $\Lambda$.  

\begin{figure}[htb]
 \parbox{\halftext}{\centerline{\includegraphics[width=7.5 cm]{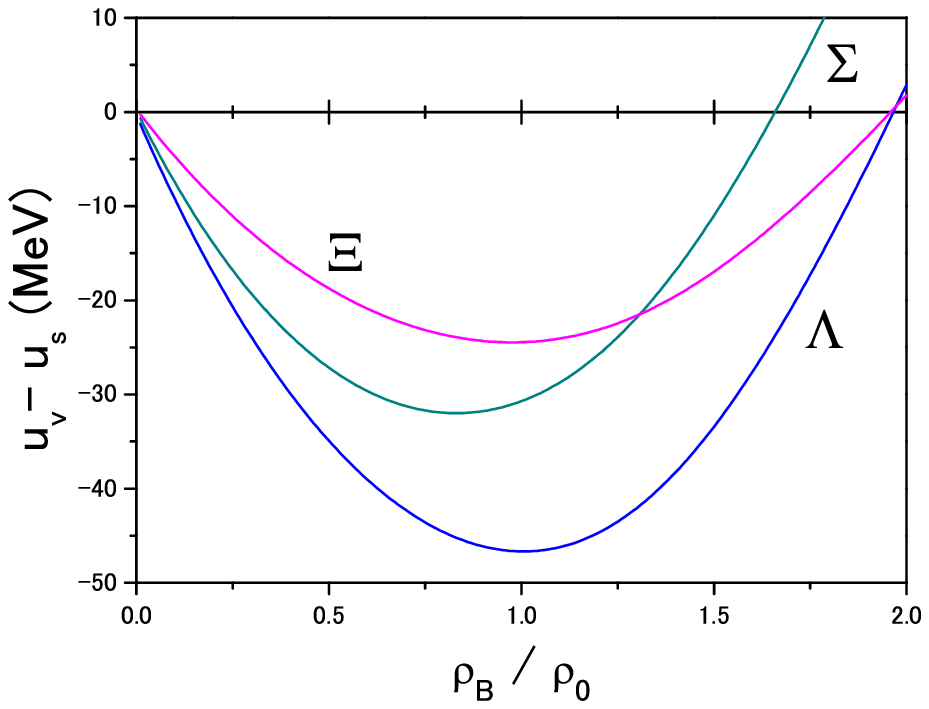}}
 \caption{Potential for the hyperon in matter (for the case 1). 
} \label{fig:pot}
}
 \hfill
 \parbox{\halftext}{\centerline{\includegraphics[width=7.5 cm]{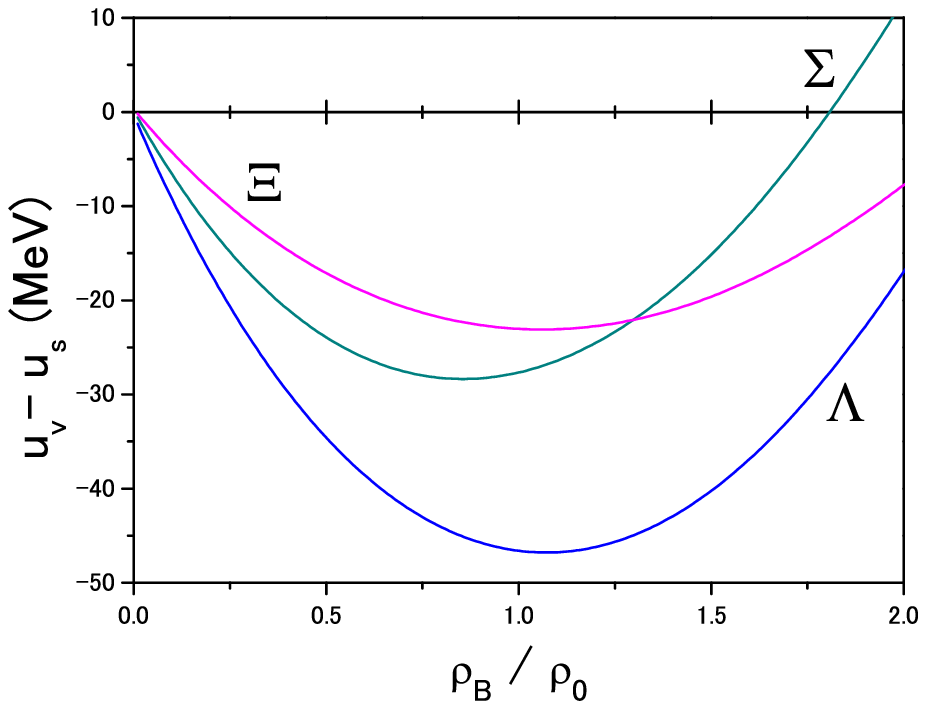}}
 \caption{Potential for the hyperon in matter (for the case 2). 
} \label{fig:potno}
}
\end{figure}
In Figs.~\ref{fig:pot} and \ref{fig:potno}, we present the potential for the hyperon, $u_v-u_s$, in matter. 
Here, $u_s (=M_Y-M_Y^*)$ is the attractive, scalar potential, and $u_v (=(n/3)g_\omega \omega$, $n$ the light quark number in $Y)$ 
is the repulsive one due to the $\omega$ exchange. 
Around $\rho_0$, the potential for the $\Lambda$ is very deep, while that for the $\Sigma$ or $\Xi$ is shallow.  
It is noticeable that, beyond $\rho_B/\rho_0 \sim 1.3$, the potential for the $\Xi$ is deeper than that for the $\Sigma$. 
The pion effect is again not large. 

Finally, we would like to add a comment concerning the present calculation. 
We consider only the pion-cloud effect on the hyperons in matter. However, the K meson also 
affects the hyperon properties, and it may be possible to perform such calculation using the SU(3) version of the CBM.\cite{cbm3}   
The HF contribution of the K-cloud may produce a further reduction of the hyperon mass, and the corrections to the $\Lambda$, $\Sigma$ 
and $\Xi$ masses are in the ratio $3 : 5 : 5$.\cite{inoue} 
However, because, as in the pion, the K meson has a finite size\cite{K} and its mass is much heavier than $m_\pi$, we can expect that 
the K-cloud effect is smaller than the pion effect. 

In summary, we have applied the chiral version of the QMC model, in which the effect of gluon and pion exchanges is included 
self-consistently, to the hyperons in a nuclear medium. The hyperfine interaction due to the OGE plays an important role in the 
in-medium baryon spectra,\cite{hyper2} while the pion-cloud effect is relatively small. 
At the quark mean-field level, the $\Lambda$ feels much more attractive force than the $\Sigma$ or $\Xi$ in matter. 
This is consistent with the experimental fact that it is not easy to discover $\Sigma$ or $\Xi$ hypernuclei. 
To draw more definite conclusions, it is necessary to make a study of finite hypernuclei including the Pauli blocking effect, 
the channel coupling, etc.\cite{hyper1,hyper2}

\section*{Acknowledgements}
We thank K. Tsushima for valuable discussions on the QMC model for hypernuclei. 
This work was supported by Academic Frontier Project (Holcs, Tokyo University of Science, 2008-2009) of MEXT. 

%

\end{document}